\documentclass[12pt,twoside,a4paper,preprint,amsmath,amssymb,nofootinbib]{revtex4}
\usepackage{graphicx}\usepackage{epsfig}
\usepackage{amsfonts}
\usepackage{cases}
 \usepackage[T2A]{fontenc}
  \usepackage[cp1251]{inputenc}
 \usepackage[russian,english]{babel}

\newcommand{\ba}{\begin{eqnarray}}
\newcommand{\ea}{\end{eqnarray}}

 \def  \Li3 { {\rm {Li}_3}  }
 \def  \li2  { { \rm {Li}_2 } }
\def  \poly4  { { \rm {Li}_4 } }
\def \Imm { \mbox{\rm Im} }
\def \Ree { \mbox{\rm Re} }

\everymath{\displaystyle}

\begin{document}
\thispagestyle{myheadings}


\title{Radiative  QED corrections to
the lepton anomaly from    four-loop vacuum polarization insertions with
three identical leptons }

\author{O.P. Solovtsova}
\email{olsol@theor.jinr.ru ;solovtsova@gstu.gomel.by}
\affiliation{Joint Institute for Nuclear Research,  \\
Joliot-Curie, 6, 141980, DUBNA, Moscow region, Russia}
\affiliation{Sukhoi  State Technical University of Gomel, \\
246029 Prospect Octiabria, 48, Gomel, Republic of Belarus}
\author{V.I. Lashkevich}
\email{lashkevich@gstu.gomel.by}
\affiliation{Sukhoi  State Technical University of Gomel, \\
246029 Prospect Octiabria, 48, Gomel, Republic of Belarus}
\author{L.P. Kaptari}
\email{kaptari@theor.jinr.ru}
\affiliation{Joint Institute for Nuclear Research,  \\
Joliot-Curie, 6, 141980, DUBNA, Moscow region, Russia}

\begin{abstract}

\vspace{0.2cm}
Analytical expressions for the tenth order electromagnetic corrections to the lepton ($L=e, ~\mu $ and $\tau$) anomaly  $a_L$ are derived explicitly for  a class of Feynman diagrams
with insertions  of the vacuum polarization operator consisting of four closed lepton loops. We
consider a particular case when one loop is formed
by the lepton $L$ of the same kind as the one under consideration, the other three
loops being formed by leptons $\ell\neq L$. The method is based on the consecutive application of dispersion relations for the polarization operator and the
 Mellin--Barnes transform for the propagators of massive particles. The result is expressed in terms of the
mass ratio $r=m_\ell/m_L$. We investigate the behaviour  of the exact analytical expressions as $r\to 0$ and $r\to \infty$ and compare them
with the corresponding asymptotic expansions known
in the literature. \end{abstract} \vspace*{6pt}

\noindent
\pacs{13.40.Em, 12.20.Ds, 14.60.Ef}


\maketitle

\section{Introduction}

As is well established, the shift of the gyromagnetic factor $g_L$ of the lepton $L$ from the
value $g_L=2$ predicted by the Dirac  theory~\cite{dirac} is entirely governed
by the  effects of  self-interactions of leptons. In the literature,  this shift is referred to
as the anomalous magnetic moment  $a_L\equiv(g_L -2)/2$.
 In spite of the extraordinary
  smallness of this anomaly, its study  is of   great importance since
the impressive accurate  measurements of  $a_L$  for  electrons~\cite{Parker:2018vye,Morel}
and muons~\cite{{E989},Fermilab2023} impose  corresponding requirements on the accuracy of  theoretical
calculations. A detailed  review of the contributions of different mechanisms to $a_L$ can be found
in e.g.,~Refs.~\cite{Jegerlehner:2017gek,review-2021}, whereas extencive discussions of the discrepancy between the
predictions of the Standard Model (SM) and experimental data are reported
in Refs.~\cite{CMD-3:2023alj,CMD-3:2023rfe,Davier-2023-1,Davier-2023-2,Greynat-2023,Masjuan-2024,Bryzgalov-2024,Volkov-2024}.
So far, the available highly accurate numerical
calculations~\cite{Laporta:2017okg,elliptic1,Volkov-2024,Volkov-2023}
 of  radiative  corrections to $a_L$  up to the eighth order were essentially based on either
  Monte Carlo integration~\cite{Volkov-2024} or    the well-known PSLQ algorithm~\cite{Bailey:1999nv}.
 Although the PSLQ  methods can assure
 up to, e.g., 1100 digits of precision~\cite{Laporta:2017okg}, numerical evaluation of the corresponding diagrams
 requires a huge amount of   computer time and, in addition,   detailed analysis of the contribution
 of individual diagrams  is hindered in such approaches.

Therefore, it is of   great interest  to be able to separate at least a subclass of Feynman
diagrams that can give results in the form  of analytically  closed  expressions and
consequently  can be used
to check the accuracy of   numerical algorithms and   investigate
the contribution of particular diagrams to   corrections from the full set of diagrams of the
a order. These are the so-called
"bubble"\ -like diagrams  consisting solely of insertions of the photon vacuum polarization operators of the desired order with the corresponding number of closed lepton loops.  In Ref.~\cite{Aguilar:2008qj} , the "bubble"\ -like diagrams were considered for the muon anomaly within the integral Mellin-Barnes representation, which allows one   to obtain analytical expressions for $a_L$ as expansions on the ratio $r=m_\ell/m_L$, where $m_L$
is the mass  of the external  lepton  $L$  and $m_\ell$  is the mass of the internal lepton  $\ell$
of the polarization operator. The approach was generalized for all kinds of
leptons $L=e,~\mu$ and $\tau$   in the whole interval of the mass ratio  $0 <\,
r\, <\, \infty$ in Ref.~\cite{Solovtsova23}, where it was argued that
  corrections of any order in $a_L$ are fully determined by diagrams with the  exchange  of only one
but massive photon. By applying the Mellin-Barnes transform to these diagrams, the general expression
for $a_L$ can be  reduced to one-dimensional integrals in the complex plane
of two Mellin momenta. The Mellin-Barnes technique  is quite popular and
widely used in the literature in multi-loop  calculations in relativistic quantum field
theories, cf.~Refs~\cite{Mellin1,Mellin2,Kotikov:2018wxe}. For the first time
this approach was implemented  in Ref.~\cite{Mellin3} as a  tool for evaluating   massive Feynman integrals.
Further developments of the approach for analytical calculations of $a_L$  can be found
 in Refs.~\cite{Friot:2005cu,Aguilar:2008qj,Solovtsova23}.

 The present paper is a generalization of the approach reported in Ref.~\cite{Aguilar:2008qj,Solovtsova23},
 but this
 time to derive  analytical expressions for corrections of the $\alpha^5$ order (where $\alpha$ is the
 fine structure constant) from diagrams with four internal lepton loops, one of which is
 formed by a lepton $L$ of the same type as the external one and the other three identical loops consisting of
 leptons $\ell$   different from $L$, $\ell\neq L$. Earlier, in Ref.~\cite{Solovtsova-Vesti},   we presented
 exact analytical expressions for   four-loop  diagrams with four identical internal loops,  whereas
 in Ref.~\cite{Solovtsova-2024} diagrams with
  two leptons of   type $L$ and two
leptons of   type $\ell$ were considered in detail.
Obviously, calculations by exact analytical formulae allow one to easily determine
not only the contributions of different diagrams with any desired accuracy  but
also to find   simpler and controllable expressions as expansions of the exact formulae at
$r\ll 1$ and $r\to \infty$. Note that
since $m_e \ll m_\mu \ll m_\tau$, actually the   limits  $r\ll 1$ and $r\to \infty$  correspond to real physical ratios of all the possible  combinations
of the known leptons $L=e,~\mu$ and $\tau$. The particular case  $r\to 0$ for corrections  from four-loop diagrams
to the  muon anomaly were reported  previosly
 in Refs.~\cite{Kataev92,Laporta94,Aguilar:2008qj}.

\section{Basic formulae}\label{glava2}
General expressions for radiative corrections from   bubble-type diagrams
with $n=p+j$ closed lepton loops, where $p$ denotes the number of loops consisting
of leptons $L$ of the same type as the external one and $j$   denotes the number of
 loops formed by leptons $\ell$ different from $L$, were considered in detail in Refs.~\cite{Aguilar:2008qj,Solovtsova23}.
 Explicitly, the  corrections $a_L(p,j)$ read as
 \begin{equation}
a_L(p,j)=\frac{\alpha}{\pi} \frac{1}{2\pi i}F_{(p,j)}
\int\limits_{c-i\infty}^{c+i\infty} dz \;
 \left( \frac{4m_{\ell}^2}{m_L^2}\right)^{-z} \Gamma(z)\Gamma(1-z)
\left(\frac{\alpha}{\pi}\right)^{p}\Omega_p(z)
\left(\frac{\alpha}{\pi}\right)^{j}R_j(z),
 \label{fin1}
\end{equation}
   \vskip 0.2cm
   \noindent
where
$F_{(p,j)} =(-1)^{p+j+1} C_{p+j}^p\;$, and $C_{p+j}^p$  are the familiar binomial coefficients;
 the variable  $c$ is an arbitrary number from the interval $a < \Ree \, z < b$  where
the integrand (\ref{fin1}) is an analytical function. The Mellin momenta
  $\Omega_p(z)$ and$R_j(z)$ are determined by the polarization operators
 $\Pi^{(L)}$ and  $\Pi^{(\ell)}$ according to
\ba && \left(\frac{\alpha}{\pi}\right)^p\Omega_p(z)=
  \int_0^1 dx \; x^{2z} (1-x)^{1-z}
  \left[ \Pi^{(L)} \left( -\frac{x^2 }{1-x}m_L^2  \right)
  \right ]^p , \label{Omp1}
  \\[0.2cm] &&
\left(\frac{\alpha}{\pi}\right)^j R_j(z)= \int_0^\infty
\frac{dy}{y}\left( \frac{4m_{\ell}^2}{y}\right)^z \frac 1\pi \Imm
\left[ \Pi^{(\ell)} (y)\right ]^j \label{rjj1}. \ea
The explicit expressions for  $\Pi^{(L,\ell)}$  in Eqs.~(\ref{Omp1}) and (\ref{rjj1})
are well known in the literature, q.v. Ref.~\cite{Lautrup:1969fr}
\ba && \Ree \; \Pi^{(L,\ell)} (y) = \left(\frac{\alpha}{\pi}\right )
\left[\frac89 - \frac{\delta^2}{3}+\delta \left(\frac12
-\frac{\delta^2}{6}\right)\ln\frac{|1-\delta|}{1+\delta}\right],
\label{reP1} \\[0.2cm] &&
\frac1\pi \Imm  \;  \Pi^{(L,\ell)}(y) = \left(\frac{\alpha}{\pi}\right
) \delta \left(\frac12 -\frac16 \delta^2\right)\theta\left
(y-4m^2_{(L,\ell)}\right), \label{imP1} \ea
where   $\delta=\sqrt{1-
4m^2_{(L,\ell)}/{y}}$.
Notice that  due to the Euclidean character of the operator $\Pi^{(L)}$ in
Eq.~(\ref{Omp1}) and due to the presence of the  $\theta$-function in Eq.~(\ref{imP1}),
the polarization operator  $\Pi^{(L)} \left( -\frac{x^2 }{1-x}m_L^2 \right)$
is purely real and does not depend on the lepton masses, i.e.,
 \ba && \Pi^{(L)} \left(
-\frac{x^2 }{1-x}m_L^2 \right)=\frac\alpha\pi \left [ \frac59
+\frac{4}{3x} - \frac{4}{3x^2}+ \left (-\frac13
+\frac{2}{x^2}-\frac{4}{3x^3}\right ) \ln(1-x)\right]. \label{lashk}
\ea

Furthermore, by a simple change of variables  $ y =
\displaystyle\frac{4m_\ell^2}{\xi}$ in Eq.~(\ref{rjj1}), it is straightforward to show that
  $ R_j(z)$ is also independent of the lepton masses. Consequently,
the only dependence of
 $a_L$ in Eq.~(\ref{fin1}) on   masses enters  through the
ratio
\begin{equation}
 r=\frac{m_\ell}{m_L}. \label{ratio}
\end{equation}
Correspondingly, in the literature it is commonly accepted
to classify the contributions to $a_L$ from different Feynman diagrams by
this ratio, emphasizing separately terms completely independent
of masses, the so-called universal contribution $A_1$
at $r=1$ and the terms $A_2(r)$ and $A_3(r_1,r_2) $ at $r\neq1$
(for details, see  Ref.~\cite{review-2021}):
\begin{equation}\label{aL}
a_{L} =  ~A_{1}\left( \frac{m_L}{m_L}\right ) + ~A_{2}\left(
\frac{m_{\ell}}{m_L}\right )  +
A_{3}\left(\frac{m_{\ell_1}}{m_L},\frac{m_{\ell_2}}{m_L}\right ).
\end{equation}
At the same time, each term in Eq.~(\ref{aL}) can be represented as
  Taylor  expansions over the fine structure constant $\alpha$  as

 \ba && A_1(m_L/m_L)= A_1^{(2)}\left(\frac\alpha\pi
\right)^1 + A_1^{(4)}\left(\frac\alpha\pi \right)^2 +
A_1^{(6)}\left(\frac\alpha\pi \right)^3 + \cdots ,
\label{A1}\\[1mm]
&& A_{2}\left({r} \right) =
A_{2}^{(4)}(r)\left(\frac\alpha\pi \right)^2 +
A_{2}^{(6)}(r)\left(\frac\alpha\pi \right)^3 +
A_{2}^{(8)}(r)\left(\frac\alpha\pi \right)^4 +A_{2}^{(10)}(r)\left(\frac\alpha\pi
\right)^5 + \cdots,
\label{A23} ~~ \\[1mm]
&& A_3\left(r_1,r_2\right)= A_3^{(6)}(r_1,r_2)\left(\frac\alpha\pi
\right)^3 + A_3^{(8)}(r_1,r_2)\left(\frac\alpha\pi \right)^4 +
A_3^{(10)}(r_1,r_2)\left(\frac\alpha\pi \right)^5 + \cdots ,~~~
\label{A4} \ea
where  $r_1= m_{\ell_1} / m_L,\ r_2= m_{\ell_2}/ m_L$, with $m_{\ell_{1,2}}$
as masses of  two internal leptons $\ell_{1,2}$ different from $L$.
The leading order contribution to $a_L$ was obtained, for the first time,
by  J.S. Schwinger~\cite{Schwinger}, $a_L= {\alpha}{/2\pi}$, which, in  our notation, corresponds to
 $A_1^{(2)}=1/2$.
The universal coefficients $A_1$ were further studied analytically in a series of publications
(see, e.g.  Refs.~\cite{Jegerlehner:2017gek,Samuel-n-bubble,latrup77,LST-2022}) for $n$ up to $n=13$.
It is worth mentioning that   $A_1^{(2n)}$ decrease for $n<7$ and, starting from
$n=7$,  increase factorially~\cite{latrup77,LST-2022}).

\begin{figure}[h!]
\centering
\includegraphics[width=0.35\textwidth]{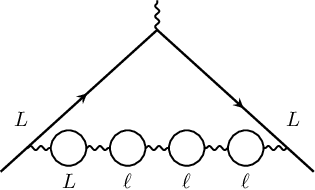}
\caption{The    considered   four-loop Feynman diagram   contributing to the tenth order radiative
corrections to the anomalous magnetic moment  $a_L$. One  loop is formed by leptons
$L$ of the same type as the external  one and the other three
   consist on leptons  $\ell\neq L$.}
   \label{Fig-1(3+1)}
\end{figure}
Recall that  the present paper is a continuation of our previous investigations of
contributions  to the mass dependent coefficients  $A_{2}^{(10)}(r)$ entering into Eq.~(\ref{A23}) of
four-loop diagrams
with    one lepton of the same type as the external
one, and  the other three   leptons $\ell\neq L$.
The corresponding diagram   is depicted in Fig.~\ref{Fig-1(3+1)}.
 In what follows, for the sake of checking the consistency of the obtained general analytical results for $A_{2}^{(10)}(r)$,
 we compare them in the limit  $A_{2}^{(10)}(  r\to 1)$ with the  analytical expressions for
 $A_{1}^{(10)} $ well-known in the literature, see e.g. Refs.~\cite{Samuel-n-bubble,Sidorov:2018pct}:
\begin{equation} \label{A10iniv}
{{A}}_{1}^{(10)} =-\frac{3 689 383}{656 100}-\frac{21 928\;\pi^4}{1
403 325}-\frac{128\;\zeta(3)}{675}+\frac{64 \;\zeta(5)}{9}
               =~4.7090571603... \times 10^{-4} \, .
\end{equation}

\section{Analytic calculations}\label{glava2}
The diagram in   Fig.~\ref{Fig-1(3+1)} implies that
$p=1$, $j=3$ and, consequently the corresponding factor $F_{(p,j)}=-4$. Then Eq.~(\ref{fin1}) for the
coefficients $A_{2}^{(10),L \ell  \ell \ell}(r) $
reads as

\begin{equation} \label{A130}
 A_{2}^{(10),L \ell  \ell \ell} (r)  = -\frac{4}{2\pi i} \int\limits_{c-i\infty}^{c+i\infty}
(4r^2)^{-z} \Gamma(z)\Gamma(1-z) \Omega_1(z)R_3(z) dz   \, ,
\end{equation}
where, for better visualization, we introduced the superscripts of $ A_{2}^{(10),L \ell  \ell \ell}$  which
directly indicate   the  type of the  considered diagram.

Straightforward calculations
of the Mellin momenta $\Omega_1(z)$ and $R_3(z)$ by Eqs. (\ref{Omp1})-(\ref{imP1})
provide

 \ba \label{Omega1}
 && \Omega_1(z)\;  = \; \frac{2^{2z}}{2\sqrt{\pi}}\bigg\{\frac{(1-2z)(1+2z)(36+54z-29z^2-34z^3+5z^4+4z^5)}{9(1-z)z(1+z)(z+2)}
\nonumber \\
 &&
 ~~~~~~~~~-2\pi(-z^2+z+1)
 \cot(\pi z) \bigg\} \Gamma(-2-z)\Gamma\left(-\frac{1}{2}+z \right),
  \label{om2}
   \ea
\begin{equation} \label{R3}
R_3(z)=\frac{\sqrt{\pi}}{864}\;\frac{\Gamma(z)}{\Gamma(\frac{11}{2}+z)}
\bigg[\frac{(z-1)\;{Y}(z)}{z(1+z)(2+z)}
 -(1+z)(35+21z+3z^2)(27\pi^2-162 \psi^{(1)}(z)
 \bigg]\,,
\end{equation}
   \vskip 0.2cm
   \noindent
where
$\psi^{(1)}(z)$  denotes the polygamma function  of the first order and, for   brevity, we introduced  the notation
$Y(z)=-3492+5256z+31831z^2+41045z^3+22650z^4+5632z^5+512z^6$.
Below, along  with the variable $r$, we widely use the variable $t=r^2$
that facilitates comparisons of our results with the corresponding expressions  well-known
 in the literature. Then in terms of $t$ the coefficients (\ref{A130}) read as

\begin{equation}
{ {\ A}}_{2}^{(10),L \ell  \ell \ell}(t) =-\frac{4}{2\pi i}
 \int\limits_{c-i\infty}^{c+i\infty} t^{-z} {\cal F}(z)  dz,
\label{Int13}
\end{equation}
where the integrand  ${\cal F} (z)$ is
   \vskip 0.2cm
   \noindent
\begin{eqnarray}
&& {\cal F} (z)
=\left[\frac{(1-2z)(1+2z)(36+54z-29z^2-34z^3+5z^4+4z^5)}{9(1-z)(1+z)z(z+2)}-2\pi(1+z-z^2)\cot
(\pi z)\right]
\nonumber \\
[0.2cm]
&& ~~ \times \bigg[ \frac{(-1+z)Y(z)}{{27}z(1+z)(2+z)}
 -(1+z)(35+21z+3z^2)\left(\pi^2-6\; \psi^{(1)}(z)\right)\bigg]
\frac{\pi^2}{Z(z)\sin^2(\pi z)}
,
\label{F13}
\end{eqnarray}
with $Z(z)$  as  the  polynomial
\begin{eqnarray}
Z(z)= z (2+z) (-1+2 z) (1+2 z) (3+2z) (5+2 z) (7+2 z) (9+2 z).
\label{Z13}
\end{eqnarray}
From Eqs.~(\ref{F13}) and (\ref{Z13}) one immediately infers that the integrand ${\cal F}(z)$  in (\ref{Int13})
is a singular function in the complex plane of the variable $z$ with numerous
poles of different orders originating from the denominator
of ${\cal F}(z)$ and  singularities due to
$\cot(\pi z)$, $\psi^{(1)}(z)$, $1/\sin^2(\pi z)$ and zeros of $ Z(z)$. Then  the
integral  (\ref{Int13}) can be calculated by the Cauchy  residue theorem  by closing
the integration contour consecutively  in the right ($t>1$) and left  ($t<1$)
semiplanes of the variable $z$.

\subsection{The left semiplane: $t< 1$}\label{glava2A}

In the left semiplane, the integrand  ${\cal F} (z)$
possesses poles at  $z=-1/2$, $-3/2$, $-5/2$, $-7/2$, $-9/2$, $0,-1,-2,-3,-4, ..., -n, \;
...$\,. The residues at  negative half-integer values  $z=-1/2$, $-3/2$, $-5/2$, $-7/2$ and $-9/2$
 are calculated directly  as residues in poles of the first order. As for the negative integers $z=-n$,
the corresponding poles are of much higher orders  (up to the sixth)   and calculations of the residues
"by hand"  are highly complicated and cumbersome.
 The high order residues are usually calculated  by means of   symbol manipulation packages such as   the
  ``Wolfram Mathematica''\ or ``Maple'' program systems with built-in libraries allowing analytical symbolic calculations. The result  for the integrals  (\ref{Int13})--(\ref{Z13}) is

\begin{eqnarray}
&& { {\ A}}_{2}^{(10),L \ell  \ell
\ell}(t<1)=P_0(t)+P_1(t)\ln(t)+P_2(t)\ln^2(t)+P_3(t)\ln^3(t)+P_4(t)\ln^4(t)
+t^{3/2} \times
\nonumber \\
 &&~~  \left(\frac{82}{45}+\frac{44
t}{7} -\frac{52762 t^2}{30375}-\frac{139001992 t^3}{343814625}-\pi
^2{g_1(t)}\right) \left[ {\rm
Li_2}\left(\frac{1-\sqrt{t}}{1+\sqrt{t}}\right) -{\rm
Li_2}\left(-\frac{1-\sqrt{t}}{1+\sqrt{t}}\right)\right]
\nonumber
\end{eqnarray}
\begin{eqnarray}
 &&~~ -g_3(t){\rm
 Li_2}\left(1-t\right)+\frac{472}{10395}(2+\pi^2)\left[{\rm
 Li_2}\left(1-t\right)-\frac{\pi^2}{6}\right]
 +
~\bigg[\frac{230}{81}+\frac{7088 t}{1225}+\frac{20714 t^2}{675}
\nonumber \\
 &&~
- \frac{28 \pi ^2 }{9}\left(\frac{1}{3}+t^2\right)
 \bigg]{\rm Li_3}(t)-\left(\frac{3716}{2835}+\frac{944}{10395 t}-\frac{16 t}{7} -\frac{4
t^2}{9}\right){\rm Li_4}(t)-\frac{40}{3}\left(1+ t^2\right)
 \nonumber \\
 &&~\times {\rm
Li_5}(t)
+
g_4(t)\Phi\left(t,3,\frac{1}{2}\right)+g_1(t)t^2\Phi\left(t,4,\frac{1}{2}\right)
 +g_2(t)\Phi\left(t,5,\frac{1}{2}\right) + \Sigma_1(t ) \, ,
\label{A13A}
\end{eqnarray}
where  ${\rm Li_n}(t)$ denote the polylogarithm functions of the order $n$ and  $\Phi\left(t,n,\alpha\right)$
is the Lerch transcendent  function: $\Phi\left(t,n,\alpha \right)=\sum_{k=0}^\infty t^k/(k+\alpha)^n)$.
In particular, at
$\alpha=1/2$  and  $\alpha=1$ the following relations hold:
 $${\sqrt{t}}\; \Phi(t,n,1/2)={2^{n-1}}\bigl[{\rm Li}_n (\sqrt{t})-{\rm
Li}_n (-\sqrt{t})\bigr], \quad {\rm Li_n}(t) = t \Phi\left(t,n,1\right).$$

The polynomial coefficients $P_n(t)$ in front of the $n$-th  power of logarithms
$\ln^n(t)$ ($n=0..4$),  include  the  dependence on  $t$, polylogarithms ${\rm Li_n}$, the
Lerch transcendent  function $\Phi\left(t,n,1\right)$
and the Riman $\zeta$-function
  ($\zeta(z)=\sum_{n=1}^\infty n^{-z}$, where, e.g.,
 $\zeta(2)=\pi^2/6$, $\zeta(4)=\pi^4/90$, etc.).
 These  coefficients are
\begin{eqnarray}
&&P_0(t)=-\frac{196219921}{25259850}-\frac{137710949
t}{5613300}+\frac{47812833977 t^2}{1375258500} +\frac{198811157
t^3}{152806500}+\frac{287301521 t^4}{1375258500}\nonumber \\
 &&
~~~~+~\pi ^2 \left(-\frac{25134061}{26943840}+\frac{3975131357
t}{1833678000}-\frac{532838 t^2}{155925}-\frac{295639
t^3}{498960}-\frac{108041
t^4}{997920}\right)\nonumber \\
 &&
~~~~ - ~\pi^2t^{3/2}\left[\frac{41}{90}+\frac{11 t}{7}-\frac{26381
t^2}{60750}-\frac{34750498 t^3}{343814625}+
\frac{1}{12}\pi^2{g_1}(t)\right] \nonumber
\\
 &&
~~~  ~-\left[\frac{476}{81}+\frac{3364 t^2}{225}-\frac{16}{9} \pi
^2\left(\frac{1}{3}+t^2\right)\right]\zeta(3) - \frac{2}{27}\pi
^4\left(\frac{131}{630}-\frac{4t}{7}+ \frac{187t^2}{150}\right) .
\nonumber
\end{eqnarray}

\begin{eqnarray}
&&P_1(t)=\frac{19410659}{5613300}-\frac{175598723
t}{11226600}-\frac{23321950727 t^2}{2750517000} +\frac{689996977
t^3}{916839000}+\frac{287301521 t^4}{2750517000}\nonumber \\
 &&~~~~
+~\pi ^2 \left(-\frac{2456747}{5987520}+\frac{17246717
t}{17463600}-\frac{9060551 t^2}{4989600} -\frac{1889
t^3}{6160}-\frac{108041 t^4}{1995840}\right)+\frac{32}{15}t ^2\zeta(3)
\nonumber \\
 &&~~~~+
\frac{53}{135}\pi^4\left(\frac{1}{3}+t^2\right)+\frac{g_4(t)}{\sqrt{t}}\left[
{\rm Li_2}\left(\frac{1-\sqrt{t}}{1+\sqrt{t}}\right)
 -{\rm Li_2}\left(-\frac{1-\sqrt{t}}{1+\sqrt{t}}\right) -\frac{\pi
^2}{4} \right]
\nonumber \\
 &&~~~~
-\left[\frac{41}{81}-\frac{3544 t}{3675}-\frac{233
t^2}{45}+\frac{2\pi^2}{9}\left(\frac{1}{3}+ t^2\right)\right]{\rm
Li_2}(1-t)
 \nonumber \\
 &&~~~~
- \left[\frac{115}{81}+\frac{3544 t}{1225}+\frac{10357 t^2}{675}
-\frac{14 \pi ^2}{9} \left(\frac{1}{3}+t^2\right)\right] {\rm Li_2}(t)
 \nonumber \\
 &&~~~~
+\left(\frac{2456}{2835}+\frac{944}{10395 t}+\frac{16 t}{105}-\frac{44
t^2}{45}\right){\rm Li_3}(t)+\frac{32}{3}\left(\frac{2}{3}+
t^2\right){\rm Li_4}(t)
\nonumber \\
 &&~~~~ -\frac{1}{2}g_4(t)\Phi\left(t,2,\frac{1}{2}\right)
 -g_1(t)t^2\Phi\left(t,3,\frac{1}{2}\right)
 -g_2(t)\Phi\left(t,4,\frac{1}{2}\right)
\,.
\nonumber
\end{eqnarray}

\begin{eqnarray}
&&P_{2}(t)=-\frac{18137003}{7484400}-\frac{2840603
t}{1663200}+\frac{573063097 t^2}{104781600} -\frac{17015189
t^3}{62868960}-\frac{2618141 t^4}{34927200}
\nonumber \\
 &&
~~~~-\pi ^2 \left(\frac{229949}{6531840}+\frac{236}{93555
t}-\frac{9617 t}{40320} +\frac{2051 t^2}{3240}+\frac{275
t^3}{10368}+\frac{19 t^4}{6912}\right) -\frac{g_1(t)}{3}t^{3/2}
\nonumber \\
 &&
~~~~\times \left[ {\rm Li_2}\left(\frac{1-\sqrt{t}}{1+\sqrt{t}}\right)
-{\rm Li_2}\left(-\frac{1-\sqrt{t}}{1+\sqrt{t}}\right)  -
\frac{\pi^2}{4}\right]- \left(\frac{1826}{8505}+\frac{944}{31185
t}+\frac{16 t}{35}-\frac{76 t^2}{135}\right)
\nonumber \\
 &&
~~~~\times \;{\rm Li_2}(t) + \left(\frac{598}{8505}+\frac{472}{31185
t}+\frac{136 t}{315}-\frac{2 t^2}{5}\right){\rm Li_2}(1-t)
-4\left(\frac{4}{9}+t^2\right){\rm Li_3}(t)
\nonumber \\
 && ~~~~
+\frac{1}{3}~{g_1(t)}t^2\Phi\left(t,2,\frac{1}{2}\right)+\frac{1}{2}~g_2(t)\Phi\left(t,3,\frac{1}{2}\right)
 \, .
 \nonumber
\end{eqnarray}
   \vskip -0.2cm
   \noindent
\begin{eqnarray} \label{P34}
&&P_{3}(t)=-\frac{1868939}{5987520}-\frac{265571
t}{1496880}-\frac{1365941 t^2}{14968800}-\frac{769337
t^3}{4490640}-\frac{151931 t^4}{5987520} + \frac{\pi
^2}{6}\left(\frac{1}{3}+{t^2}\right)
\nonumber \\
 &&
 ~~~~~
  +\frac{g_2(t)}{12\sqrt{t}}
\left[ {\rm Li_2}\left(\frac{1-\sqrt{t}}{1+\sqrt{t}}\right) -{\rm
Li_2}\left(-\frac{1-\sqrt{t}}{1+\sqrt{t}}\right) -\frac{\pi ^2}{4}
\right] +\frac{7}{9} \left(\frac{1}{3}+t^2\right){\rm Li_2}(t)
\nonumber \\
 &&
 ~~~~~
-\frac{1}{9}\left(\frac{1}{3}+t^2\right){\rm Li_2}(1-t)
-\frac{1}{8}~g_2(t)\Phi\left(t,2,\frac{1}{2}\right),
\nonumber \\[0.2cm]
 &&
P_4(t)= -\frac{37}{1536}+\frac{49 t}{6912}-\frac{313
t^2}{2160}-\frac{275 t^3}{20736}-\frac{19 t^4}{13824}, \quad
\nonumber
\end{eqnarray}
where for brevity the following notation is introduced:
\begin{eqnarray}
&& g_1(t)=\frac{41}{180}+\frac{99 t}{140}-\frac{713
t^2}{3780}-\frac{5381 t^3}{124740},
\nonumber \\
&&g_2(t)=\frac{37}{128}+\frac{101 t}{384}+\frac{451 t^2}{192}+\frac{31
t^3}{64}+\frac{59 t^4}{384}+\frac{19 t^5}{1152}\, ,
\nonumber \\
&&g_3(t)=\frac{2593}{729}+\frac{20019064 t}{3472875}-\frac{63253
t^2}{10125}-\pi ^2 \left(\frac{598}{2835}+\frac{136 t}{105}-\frac{6
t^2}{5}\right),
\nonumber \\
&&g_4(t)=\frac{22927}{21600}-\frac{451 t^2}{48}-\frac{155
t^3}{72}-\frac{15281 t^4}{21600} -\frac{61351 t^5}{793800} +\frac{~
\pi ^2}{6}g_2(t),
\nonumber \\
&&g_5(t)=\frac{64}{675}+\frac{101 t}{96}+\frac{2255
t^2}{216}+\frac{8029 t^3}{3600} +\frac{190511
t^4}{264600}+\frac{2231911 t^5}{28576800}+\frac{\pi ^2}{6}g_2(t),
\nonumber \\
&&g_6(t)=\frac{164}{81}+\frac{814 t}{125}-\frac{18418216
t^2}{10418625} -\frac{114927398 t^3}{281302875}+\frac{\pi
^2}{3}g_1(t). \nonumber
 \label{g13}
\end{eqnarray}
Notice that the functions  $g_5(t)$ and  $g_6(t)$ are also present
in the expressions for   ${ {\ A}}_{2}^{(10),L \ell  \ell \ell}(t)$ in the
right semiplane of the variable $z$, see below.
Eventually, the sum  $\Sigma_1(t )$ is
\begin{eqnarray}
~~~ \Sigma_1(t )&= &\frac{8}{3}\; \sum_{n=3}^{\infty}\bigg\{ \left[
Q_1(-n)+Q_2(-n)\ln(t)-Q_3(-n)\ln^2(t) \right]\psi^{(1)}(n)
 \nonumber \\[-0.1cm]
&& +\big[Q_2(-n)-2Q_3(-n)\ln(t) \big] \psi^{(2)}(n)
-Q_3(-n)\psi^{(3)}(n) \bigg\} {t^{n}} \, ,
 \label{Sum-13-A}
\end{eqnarray}
where  $\psi^{(k)}(n)$ denotes the polygamma function of the order $k$  of
integer arguments $n$;  $ Q_{i=1,2,3}$ are defined as
\begin{eqnarray}
&&  Q_1(n)=6751269000+40950554400 n+19613108340
n^2-383309558856 n^3  \nonumber \\
 &&
~~ -826774862139 n^4
 +924941026044 n^5+5538290450480 n^6 + 6540701065768
n^7  \nonumber \\
 &&
~~ -2601992148019 n^8 -13947133685004 n^9-12413008740734 n^{10}+1058973953952
n^{11}\nonumber \\
 &&
~~ +10775553931592 n^{12} +9161255067136 n^{13}+2818170076544
n^{14}-844629934848 n^{15}
\nonumber \\
 &&
~~ -1166425861376 n^{16} -492797758464 n^{17}-92025138688 n^{18}+4832247808 n^{19}
\nonumber \\
 &&
~~ +7132129280 n^{20}+1858043904 n^{21} +253444096 n^{22}+18677760 n^{23}
\nonumber \\
 &&
~~
+589824 n^{24}
 / \left[(n^2-1)Z^3(n)\right] ,
 \nonumber
\end{eqnarray}
   \vskip -0.3cm
   \noindent
\begin{eqnarray}
&& Q_2(n)= 2381400+9315810 n-9561069 n^2-82903809 n^3-100877126
n^4+40692827 n^5\nonumber \\
 &&
~~~~~~~ +164693675 n^6+103792048 n^7-10480368 n^8-38257440 n^9-14786656
n^{10} \nonumber \\
 &&
~~~~~~~ +88768 n^{11}+1590016 n^{12}+483584 n^{13}+62208 n^{14}+3072 n^{15}
/ \left[(n^2-1)Z^2(n)\right],
\nonumber\\
&& Q_3(n)= 9 (n^2-n-1)(3n^2+21n+35)
/Z(n) .
\nonumber
\end{eqnarray}

Thus,   the resulting  analytical expression (\ref{A13A}) for
  $A_{L}^{L \ell  \ell  \ell}(t<1)$ turns out to be  quite lengthy containing
a number of the above  mentioned special functions and high order polynomials  of  the variable $t=r^2$.

\subsection{The right semiplane: $t>1$}\label{glava2B}
In the same manner, we calculate  and sum    up all the residues in the right semiplane of $z$.
The result is
\begin{eqnarray}
&&\quad { {\ A}}_{2}^{(10),L \ell \ell \ell}(t>1) = \; D_0(t)+
D_1(t)\ln(t) + D_2(t)\ln^2(t)+g_6(t) t^{3/2}
 \left[ {\rm Li_2}\left(\frac{\sqrt{t}-1}{\sqrt{t}+1}\right)  \right. \nonumber \\
 &&~~~~
 \left. -{\rm
Li_2}\left(-\frac{\sqrt{t}-1}{\sqrt{t}+1}\right)
 -\frac{\pi^2}{4} \right] + \bigg[\frac{658010786}{281302875}+\frac{29015464
t}{3472875}-\frac{76078 t^2}{10125}+\pi ^2 \left(\frac{598}{8505}
 \right.
\nonumber \\
 &&~~~~
\left.+\frac{472}{31185 t} +\frac{136 t}{315}-\frac{2 t^2}{5}\right)\bigg] {\rm
Li_2}\left(1-\frac{1}{t}
\right)+\frac{4g_5(t)}{\sqrt{t}}
\left[  {\rm
Li_3}\left(\frac{1}{\sqrt{t}}\right) - {\rm
Li_3}\left(-\frac{1}{\sqrt{t}}\right) \right]
\nonumber\\
&&~~~~
+\left[\frac{1486136}{297675} +
\frac{7088 t}{1225} +
\frac{22064 t^2}{675}+\frac{4}{3}\pi^2 \left(\frac{1}{3}+t^2\right)
\right]{\rm
Li_3}\left(\frac{1}{t}\right)-\left(\frac{1552}{945}+\frac{512
t}{105}
\right.
\nonumber \\
 &&~~~~
\left.
-\frac{128 t^2}{45}\right){\rm
Li_4}\left(\frac{1}{t}\right)+\Sigma_2(t),
\label{A13B}
\end{eqnarray}
where
\begin{eqnarray}
&&D_{0}(t)= \frac{8741299421}{723350250}+\frac{40331092741
t}{1125211500}+\frac{153650032027 t^2}{28130287500}+\frac{95289439723
t^3}{10126903500}
\nonumber \\[0.1cm]
 &&~~~~~~~
+\frac{1622471149 t^4}{1125211500}+\pi^2 \bigg(\frac{63836886989}{297055836000}+\frac{4838}{93555
t}-\frac{101990719 t}{611226000}+\frac{20342647
t^2}{14033250}
\nonumber \\
 &&~~~~~~~
+\frac{4507981 t^3}{13471920}+\frac{151931
t^4}{2993760}\bigg) - \;\frac{\pi ^4}{45}
\left(\frac{299}{567}+\frac{236}{2079 t}+\frac{68 t}{21}-3 t^2\right)
, \nonumber\\[0.2cm]
&& D_{1}(t)=\frac{5884987}{1148175}+\frac{18673392163
t}{2250423000}+\frac{38398840507 t^2}{11252115000}
+\frac{32532326467 t^3}{6751269000}+\frac{1622471149
t^4}{2250423000}\nonumber \\
 &&~~~~~~
+ \; \pi^2\left(\frac{1868939}{5987520}+\frac{265571
t}{1496880}+\frac{1839029 t^2}{14968800}+\frac{769337 t^3}{4490640}
+\frac{151931 t^4}{5987520}\right)\nonumber \\
 &&~~~~~
+\frac{g_5(t)}{\sqrt{t}} \bigg[ {\rm
Li_2}\left(\frac{\sqrt{t}-1}{\sqrt{t}+1}\right)-{\rm
Li_2}\left(-\frac{\sqrt{t}-1}{\sqrt{t}+1}\right)-\frac{\pi ^2}{4}
 - {\rm Li_2} \left(
\frac{1}{\sqrt{t}} \right)+ {\rm Li_2}
\left(-\frac{1}{\sqrt{t}}\right) \bigg]
\nonumber
\\
 &&~~~~~
-\left[\frac{743068}{893025}+\frac{3544 t}{3675}+\frac{248
t^2}{45}+\frac{2}{9} \pi^2 \left(\frac{1}{3}+t^2\right) \right]
\left[{\rm Li_2}\left(1-\frac{1}{t}\right)-\frac{\pi ^2}{6}\right]
+\left[\frac{743068}{297675} \right.
\nonumber \\
 &&~~~~
 \left.
+ \; \frac{3544 t}{1225}+\frac{11032 t^2}{675}+\frac{2}{3} \pi^2
\left(\frac{1}{3}+t^2\right) \right] {\rm
Li_2}\left(\frac{1}{t}\right) -\left(\frac{776}{945}+\frac{256
t}{105}-\frac{64 t^2}{45}\right){\rm Li_3}\left(\frac{1}{t}\right) ,
 \nonumber
\end{eqnarray}
   \vskip -0.2cm
   \noindent
\begin{eqnarray}
&&D_{2}(t)=\frac{482938648}{281302875}+\frac{5591300261
t}{1000188000}+\frac{355194901 t^2}{142884000} +\frac{2558299
t^3}{3429216}+\frac{2231911 t^4}{28576800}\nonumber \\
 &&~~~~~~~~
+\;\pi ^2 \left(\frac{181439}{2177280}+\frac{236}{31185
t}+\frac{24397 t}{120960}+\frac{97 t^2}{1080}+\frac{275 t^3}{10368}
+\frac{19 t^4}{6912}\right)\nonumber \\
 &&~~~~~~~~+\left(-\frac{388}{2835}-\frac{128
t}{315}+\frac{32 t^2}{135}\right){\rm
Li_2}\left(\frac{1}{t}\right),\nonumber
\end{eqnarray}
 and
\begin{eqnarray}
\Sigma_2(t) &=&  \frac{8}{3}\; \sum_{n=2}^{\infty}\bigg\{ \left[
Q_1(n)+Q_2(n)\ln(t)-Q_3(n)\ln^2(t) \right]\psi^{(1)}(n)
 \nonumber \\
&& ~~~~ -\big[Q_2(n)-2Q_3(n)\ln(t)\big] \psi^{(2)}(n) -Q_3(n)\psi^{(3)}(n)
\bigg\} \frac{1}{t^{n}} \,.
 \label{Sum-13-B}
\end{eqnarray}
In Eq.~(\ref{Sum-13-B})б еру  notation for the polynomials $Q_i(n)$ is the same as used for  $\Sigma_1(t)$,
q.v. Eq.~(\ref{Sum-13-A}).

\section{Numerical results}\label{numerical}

The above  expressions (\ref{A13A}) - (\ref{Sum-13-B})
determine explicitly the coefficients $A_{2}^{(10), L \ell \ell\ell}(t)$
and allow one to perform numerical calculations with any predetermined accuracy. Moreover,
the accuracy of calculations is only limited by our knowledge
of the experimentally measured fundamental constants, viz. the fine structure constant
$\alpha$ and the lepton masses $m_\ell$ and $m_L$.
ex
\begin{figure}[h]
\includegraphics[width=0.6\textwidth]{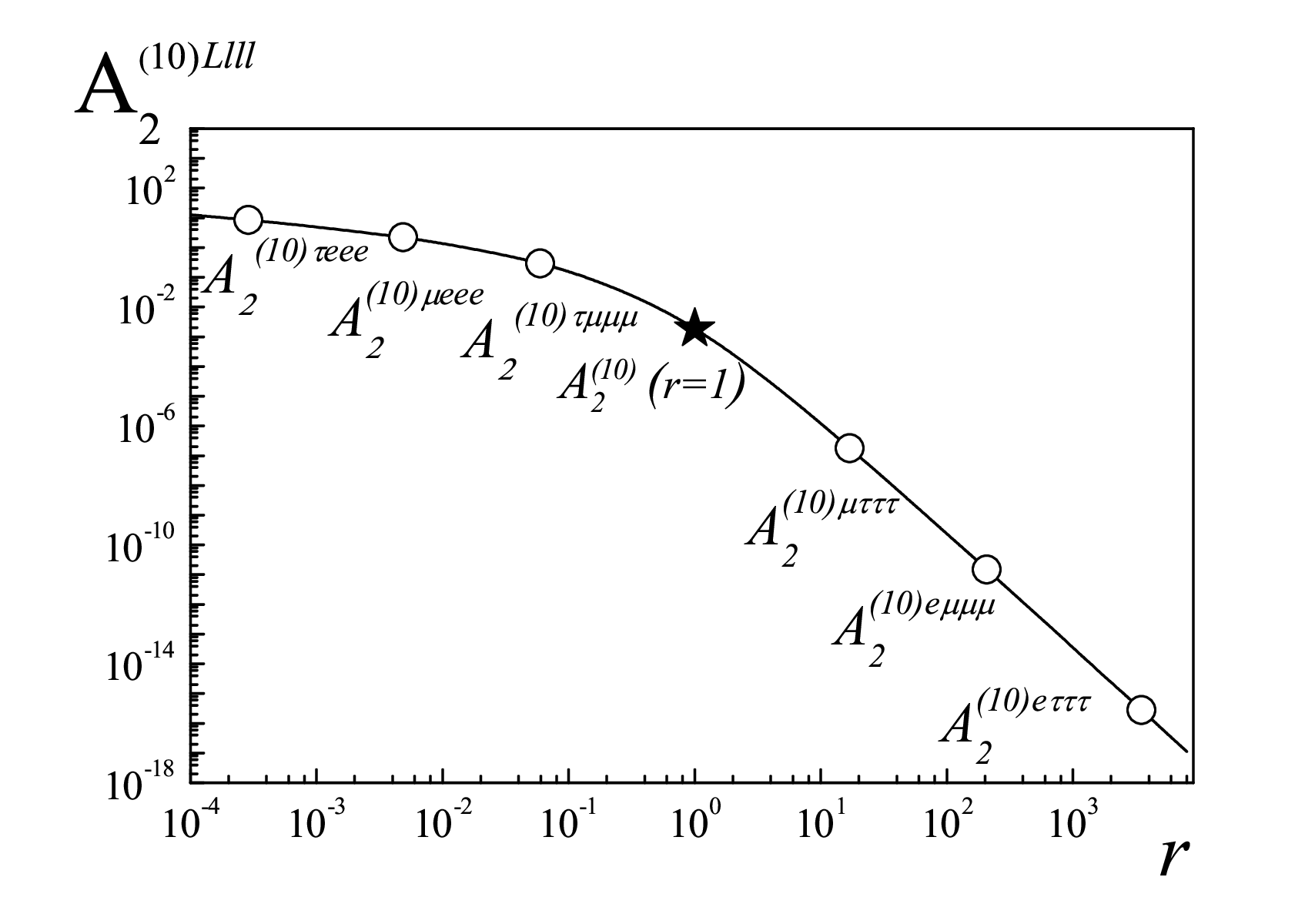
}\hspace*{1cm}
\caption{Behaviour of the coefficient   ${ {\ A}}_{2}^{(10), L \ell \ell  \ell}(r)$
 as a function of the lepton mass ratio   $r=m_\ell/m_L$. The open circles indicate  on
 physical values of ${ {\ A}}_{2}^{(10), L \ell \ell  \ell}(r) $ for  the real
   existing leptons. The star denotes the universal value at
  $r=1$.}
\label{A1013AB}
\end{figure}
Figure~\ref{A1013AB} represents the numerical calculations of
${ {\ A}}_{2}^{(10), L \ell \ell  \ell}(r)$  by Eqs.~ (\ref{A13A}) - (\ref{Sum-13-B}) (solid line) as a function
of the ratio $r=m_\ell/m_L$  \footnote{For more transparent treatments of the results, we again use the variable $r$ instead of  $t=r^2$}
in a large interval   $0 < \, r\, <\, \infty$. For real existing
leptons, the coefficients ${ {\ A}}_{2}^{(10), L \ell \ell  \ell}(r)$  are additionally
emphasized by open circles and labeled explicitly. The universal value at   $r=1$ is represented by a star.
It can be seen that the main contribution to ${ {\ A}}_{2}^{(10), L \ell  \ell \ell}(r)$ comes from
the diagrams with three lightest leptons (electrons).
  With increasing   mass of the loop leptons, the coefficient
    ${\ A}_2^{(10), L \ell  \ell \ell}(r)$ drops down sharply  by about 18 orders of magnitudes, drops down, i.e.,
  for heavier leptons  in the loop, $\ell= \mu$ and $\ell=\tau$,
  the coefficients ${{\ A}}_{2}^{(10),  e  \mu \mu \mu }$ , ${{\ A}}_{2}^{(10), \mu \tau \tau \tau }$
  and  ${{\ A}}_{2}^{(10),  e  \tau \tau  \tau}$, for which $r\gg 1 $,
become extremely small.  However, in view   of the high accuracy of
 measurements,  these coefficient are rather important
  in estimations of the theoretical predictions and   reliability of the approach.
\begin{figure}[hbt]
\includegraphics[width=0.49\textwidth]{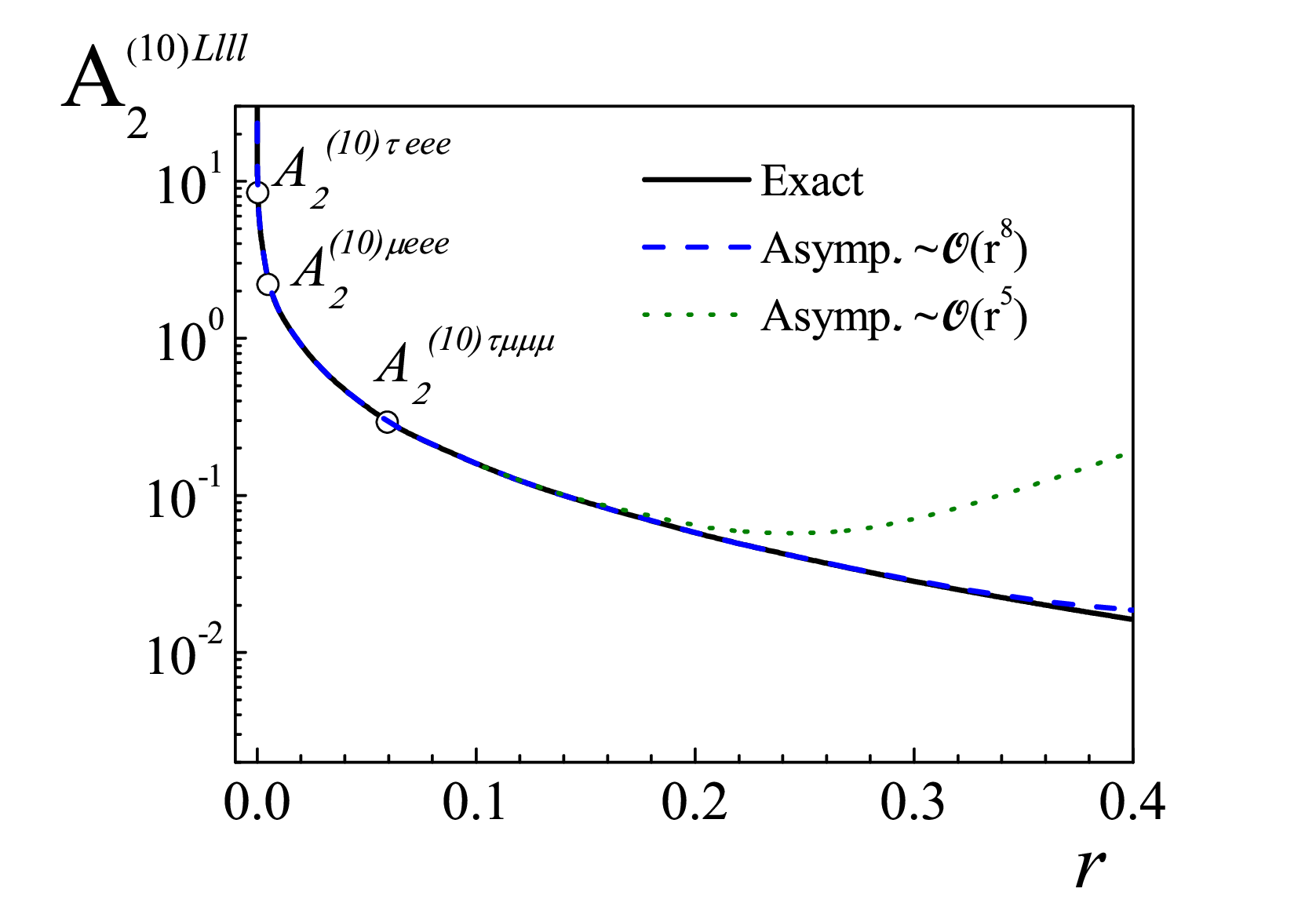}
\includegraphics[width=0.49\textwidth]{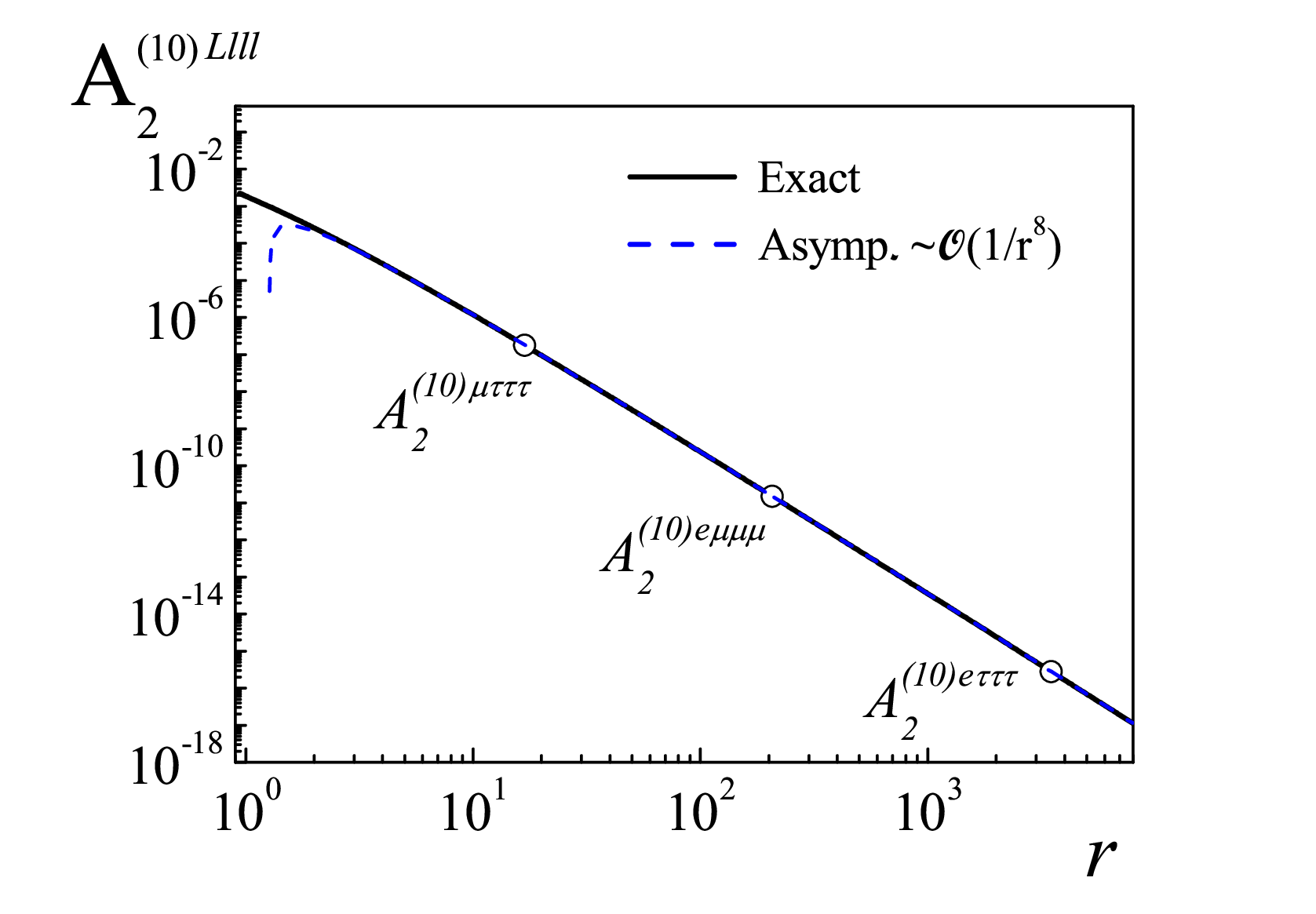}
\caption{Comparison of the exact results with the asymptotic
expansions.
 Left panel: the solid curve depicts the  calculations by
  Eq.~(\ref{A13A}), the dotted curve represents the calculations by asymptotic expansion
at $r<1$ keeping   terms up to  ${\cal O}(r^{5})$ that  correspond to expansion
reported in Ref.~\cite{Aguilar:2008qj},  the dot-dashed curve
is the results of  calculations by ~Eq.~(\ref{asym-13-A}) keeping
terms up to  ${\cal O}(r^8)$.  Right panel: the solid curve is the result of calculations
by exact formula Eq.~(\ref{A13B}),  the  dashed curve  -- asymptotic calculations at  $r>1$  with terms up to ${\cal O}(1/r^{8})$} \label{RisAsymp}
\end{figure}
Notice that for all the possible combinations of leptons in  the interval $r<1$,
the maximum value is   $r_{max}=m_\mu/m_\tau \simeq 5.95 10^{-2}\ll 1$, whereas in  the interval $r>1$
the minimum possible value of $r$ is
 $r_{min}=m_\tau/m_\mu \simeq  16.8 \gg 1$ (see Ref.~\cite{CODATA}).
This indicates that in reality  the physical values of $r$ are either in the intervals
  $r  < 0. 06\ll 1$  or $r > 16\gg 1$.
  This circumstance serves as a hint  that the complicated expressions, Eqs.~(\ref{A13A})-(\ref{g13}) at $r<1$ and
Eqs.~(\ref{A13B})-(\ref{Sum-13-B}) at $r>1$ can be replaced by   simpler Taylor expansions
as $r\to 0$ and $r\to\infty$, respectively. Then, in the Taylor series on can keep as many terms as necessary to assure the desired accuracy in numerical calculations.

The corresponding expansion of Eqs.~(\ref{A13A})-(\ref{g13}) at small
$t=r^2\ll 1$  up to terms  $\sim{\cal {O}}(t^{4})$, reads as
\begin{eqnarray}
&&A_{2,asymp.}^{(10), L \ell \ell
\ell}(t)  \underset{t \ll 1}=  \;
\frac{64244}{6561}+\frac{8}{405}\pi^4-\frac{2593}{2187}\pi^2-\frac{476}{81}\zeta(3)+\left(-\frac{119}{243}+\frac{4}{81}\pi^2\right)\ln^3(t)  \nonumber \\
&&-\left(\frac{61}{243}-\frac{2}{81}\pi^2\right)\ln^2(t)-\left(\frac{7627}{1458}+\frac{52}{81}\pi^2-\frac{16}{135}\pi^4\right)\ln (t) - \left[\frac{17525}{729}-\frac{76}{243}\pi^2-\frac{32}{135}\pi^4\right.\nonumber  \\
&&\left.+\left(\frac{230}{81}-\frac{8}{27}\pi^2\right)\ln^2(t) +\left(\frac{454}{27}-\frac{16}{9}\pi^2\right)\ln (t) \right]t + \left[\frac{675818203}{13668750}-\frac{592}{2025}\pi^4-\frac{62}{243}\pi^2\right.\nonumber
\\
&&\left.+\frac{16}{9}\pi^2\zeta(3)-\frac{3364}{225}\zeta(3)+\frac{2}{45}\ln^4(t)-\left(\frac{2459}{2025}-
\frac{4}{27}\pi^2\right)\ln^3(t)+\left(\frac{19268}{3375}-\frac{26}{81}\pi^2\right)\ln^2(t)\right.\nonumber \\
&&\left.+\left(-\frac{8929444}{455625}-\frac{98}{81}\pi^2+\frac{16}{45}\pi^4+\frac{32}{15}\zeta(3)\right)\ln
(t)\right]t^2+\left[\frac{15420978231787}{1837845450000}-\frac{125}{243}\pi^2\right.\nonumber \\
&&\left. + \frac{88}{4725}\pi^4-\frac{23192}{11025}\zeta(3)-\frac{22}{945}\ln^4(t)-\frac{5798}{33075}
\ln^3(t)+\frac{978421}{6945750}\ln^2(t)-\left(\frac{14067530059}{4375822500}  \right.\right.\nonumber \\
&&\left.\left.+\frac{4}{81}\pi^2+\frac{352}{315}\zeta(3)\right)\ln (t)
\right]t^3
-\frac{41}{540}\pi^4\;t^{3/2}-\frac{33}{140}\pi^4\;t^{5/2}+\frac{713}{11340}\pi^4\;t^{7/2}+
{\cal {O}}\left(t^4\right).
 \label{asym-13-A}
\end{eqnarray}
Expression  (\ref{asym-13-A}) is in   full agreement with the expansions reported before in
Refs.~\cite{Laporta94} and \cite{Aguilar:2008qj}, which were derived with accuracies
 $\sim{\cal {O}}(t^{1/2})$ and  $\sim{\cal {O}}(t^{5/2})$, respectively.

 Further,  from  Eqs.~(\ref{A13B})- (\ref{Sum-13-B})  it follows that at large  $t$ one has
\begin{eqnarray}\label{4-loop>1}
&& A_{2,asymp.}^{(10), L \ell \ell
\ell}(t) \underset{t \gg 1} = \; \left[\frac{6007320707}{328706428050}
-\frac{89}{675675}\pi^4-\frac{8651128}{2029052025}\zeta(3)\right. \nonumber \\
&&\left.+\left(\frac{87709}{7297290}-\frac{356}{45045}\zeta(3)\right)\ln
t\right]\frac{1}{t^2}+\left[\frac{31956039083}{61632455259375}-\frac{8}{81081}\pi^4-\frac{4}{3375}\ln^2
t\right. \nonumber \\
&&\left.+\frac{24687032}{6087156075}\zeta(3)+\left(\frac{15610069}{1368241875}-\frac{160}{27027}\zeta(3)\right)\ln
t\right]\frac{1}{t^3}+ {\cal {O}}\left(\frac{1}{t^4}\right)\,.
 \label{asym-13-B}
\end{eqnarray}
As mentioned, the asymptotic expansions (\ref{asym-13-A})-(\ref{asym-13-B}) entirely cover all
the physical values of the variable $r$ relevant to the existing leptons.
In Fig.~\ref{RisAsymp} we present the comparison of our exact results with the expansions
 as  $r\to 0$, left panel,  and $r\to\infty $, right panel. It is seen that the approximate
 calculations visually coincide with the exact result for $(0<r<0.1)$ and $(2<r< \infty)$. This
 implies that the approximate expressions (\ref{asym-13-A}) and (\ref{asym-13-B}) can be
 quite safely employed not only for qualitative estimations but also for quantitative analysis of the coefficients  $A_2^{(10)L\ell\ell\ell}(r)$ for
  any lepton $L=$~$e$, $\mu$ or $\tau$ with insertions of any type $e$, $\mu$ and $\tau$ of leptons.
  The reliability  of the  asymptotic expansions and the limits of their applicability  can be inferred
  if one  defines the relative deviation  $\varepsilon_L(r)$ from the exact results as

\begin{equation}
\varepsilon_L(r)=\frac{|A_{2,asymp.}^{(10), L \ell \ell \ell}(r)  - A_{2,exact}^{(10), L \ell \ell \ell}(r)|}
{A_{2, exact}^{(10), L  \ell \ell \ell}(r)}.
\nonumber
\end{equation}
Then  for, e.g.,  the muon anomaly,
 the  maximum contribution  comes from the four-loop vacuum polarization operator with
insertions  of   three electrons.
In this case  ($L=\mu$ and $\ell=e $), the electron-muon mass ratio is
$r_e$=0.00483633169(11) (see Ref.~\cite{CODATA}) and calculations by Eq.~(\ref{A13A}) provide
 $ A_{2,exact}^{(10), \mu e  e   e }(r_e)=2.2033273165555885939...$,
 whereas calculations by the approximate formula~(\ref{asym-13-A}) keeping terms up to ${\cal O}(r^5)$
 (this approximation corresponds to the one reported in  Ref.~\cite{Aguilar:2008qj}) result in
$ A_{2,asymp}^{(10), \mu e  e   e }\big(  \sim{\cal O}(r^5) \big)=2.20332731662.$
In  terms of the relative errors $\varepsilon_\mu(r_e)$ this corresponds to $\varepsilon_\mu(r_e) \big(  \sim{\cal O}(r^5) \big) \approx 2.8 \times 10^{-11}$. The relative errors rapidly decrease if in Eq.~(\ref{asym-13-A}) one keeps terms up to
${\cal O}(r^8)$. In this case
$A_{2,asymp}^{(10), \mu e  e   e }\big(  \sim{\cal O}(r^8) \big)=2.20332731655558856$ and,
and consequently,  $\varepsilon_\mu(r_e)(\sim{\cal O}(r^8) \big )\approx
1.4 \times 10^{-17}$. Hence, in calculations of the electron corrections to the muon anomaly it is quite sufficient
to restrict oneself to terms
  $ \sim {\cal O}(r^5)$ which assure  accuracies higher than the experimental  errors
 $\Delta r$   related to the measured~\cite{CODATA} ratio of electron  to muon masses  $ \Delta r\sim 10^{-10}$.
To estimate how far from $r\to 0$ one can apply the approximate formula, Eq.~(\ref{asym-13-A}),
we compare the exact results with the expansions
   $ \sim {\cal O}( r^5)$ and $ \sim{\cal O}(r^8)$ at $r=0.1$. We obtained that keeping
   terms  $\sim  {\cal O}(r^5)$,    Eq.~(\ref{asym-13-A})  assures only three significant digits
  while keeping terms   $ \sim{\cal O}(r^8)$, the approximate formula provides much more accurate results, namely up
  to seven significant digits in the interval $(0<\, r\, < 0.1)$. This accuracy is above  the experimental measurements in this  interval. An analogous situation occurs also in the region  $r>2$.

\section{Summary  }\label{summ}
In this paper we have presented, for the first time,
exact analytical expressions for the tenth order corrections to the
anomalous magnetic moments of leptons $e$,~$\mu$~and~$\tau$
induced by   the Feynman diagrams with insertions of the vacuum polarization operator
with four closed lepton loops. We
considered the particular case when one loop is formed by the lepton $L$ of the
same kind as the external one and the other three loops consist of  leptons $\ell\neq L$.
  The approach essentially relies on  the dispersion relations and the Mellin-Barnes
  transform   for the propagators of massive photons.
  This method allows one ]to derive explicitly the
  corresponding tenth order corrections $a_L$ as functions of the  ratio $r=m_\ell/m_L$ of the mass of the internal $\ell$
to the mass of the external $L$ leptons    in the whole interval
 ($0<r<\infty$). We presented the results of numerical calculations of $a_L(r)$ and analysed
 the contribution of different types of insertions of the vacuum polarization
 operator. It is argued that for every type of
 leptons the main contribution to $a_L(r)$ is provided by insertions of the polarization operator with
 three lightest leptons, i.e. for $r<1$.  The resulting expressions turn out to be extremely complicated and cumbersome. However, since in reality for physically existing leptons one has either
 $r\ll 1$, or $r\gg 1$, it is appropriate to replace the exact
 expressions by their asymptotic expansions, which are much simpler and more convenient
 for numerical calculations. The corresponding expansions at $r\ll 1$ and $r\gg 1$ were
 derived and the limits of their applicability were investigated. We argued that the asymptotic
 expansions work quite well in the intervals
  ($0\,< \,r\, <\, 0.1$) and  ($2\,< \,r\, <\,\infty$).

\section{Acknowledgments}
This work was supported in part
by  a grant under the Belarus-JINR scientific collaboration.


\end{document}